\documentclass[runningheads]{llncs}
\usepackage{graphicx}
\usepackage[T1]{fontenc}    
\usepackage{afterpage}      
\usepackage{csquotes}
\usepackage{float}
\floatstyle{ruled}
\newfloat{program}{t}{lop}
\floatname{program}{Program}
\usepackage{fancyvrb}
\usepackage{parcolumns}
\usepackage{wrapfig}
\usepackage{lipsum}
\usepackage{hyphenat}
\usepackage{alltt,amsmath,amssymb}
\usepackage{cite}

\usepackage{color}

\newcommand{\psl}{\texttt{PSL}}
\newcommand{\pgt}{\texttt{PGT}}
\newcommand{\pamper}{\texttt{PaMpeR}}
\newcommand{\meloid}{\texttt{MeLoId}}
\newcommand{\lifter}{\texttt{LiFtEr}}
\newcommand{\etal}{\textit{et al.}}
\newcommand{\numbvar}{\texttt{number}}
\newcommand{\rulevar}{\texttt{rule}}
\newcommand{\trmvar}{\texttt{term}}
\newcommand{\occvar}{\texttt{term\_occurrence}}

\newcommand{\execMdl}{\texttt{exec\_append\_model\_prf}}
\newcommand{\execAlt}{\texttt{exec\_append\_alt\_prf}}

\begin{document}
%
\title{\lifter{}: Language to Encode Induction Heuristics \\for Isabelle/HOL
\thanks{
We thank 
Ekaterina Komendantskaya,
Josef Urban, 
Kenji Miyamoto,
and anonymous reviewers for APLAS2019
for their valuable comments on an early draft of this paper.
This work was supported by the European Regional Development Fund under the project 
AI \& Reasoning (reg. no.CZ.02.1.01/0.0/0.0/15\_003/0000466).
}
}
%
%
\author{Yutaka Nagashima\inst{1}\inst{2}
\orcidID{0000-0001-6693-5325}}
\authorrunning{Y. Nagashima}
%
\institute{
Czech Technical University in Prague,
Prague, Czech Republic
\email{Yutaka.Nagashima@cvut.cz}
\and
University of Innsbruck, Innsbruck, Austria}

\maketitle              
\begin{abstract}
Proof assistants, such as Isabelle/HOL, 
offer tools to facilitate inductive theorem proving. 
Isabelle experts know how to use these tools effectively; 
however, there is a little tool support for transferring
this expert knowledge to a wider user audience.
To address this problem,
we present our domain-specific language, \lifter{}. 
\lifter{} allows experienced Isabelle users to encode their induction heuristics
in a style independent of any problem domain.
\lifter{}'s interpreter mechanically checks 
if a given application of induction tool matches the heuristics, 
thus automating the knowledge transfer loop.
\end{abstract}

\section{Introduction}
Consider the following reverse functions, \verb|rev| and \verb|itrev|, presented in 
a tutorial of Isabelle/HOL \cite{isabelle}:

\begin{verbatim}
primrec rev::"'a list =>'a list" where
  "rev  []      = []"
| "rev (x # xs) = rev xs @ [x]"

fun itrev::"'a list =>'a list =>'a list" where
  "itrev  []    ys = ys"
| "itrev (x#xs) ys = itrev xs (x#ys)"
\end{verbatim}
\noindent
where \verb|#| is the list constructor, 
and \verb|@| appends two lists.
How do you prove the following lemma?
\begin{verbatim}
lemma "itrev xs ys = rev xs @ ys"    
\end{verbatim}
\noindent
Since both \verb|rev| and \verb|itrev| are defined recursively,
it is natural to imagine that we can handle this problem by applying induction.
But how do you apply induction and why?
What induction heuristics do you use?
In which language do you describe those heuristics?

Modern proof assistants (PAs), such as Isabelle/HOL \cite{isabelle},
are forming the basis of trustworthy software.
Klein \etal{}, for example, verified the correctness of the seL4 micro-kernel in Isabelle/HOL \cite{sel4},
whereas Leroy developed a certifying C compiler, CompCert, using Coq \cite{compcert}.
Despite the growing number of such complete formal verification projects,
the limited progress in proof automation still keeps the cost of proof development high,
thus preventing the wide spread adoption of complete formal verification.

A noteworthy approach in proof automation for proof assistants is hammer tools \cite{hammering}.
Sledgehammer, for example, exports proof goals in Isabelle/HOL to various external automated theorem provers (ATPs)
to exploit the state-of-the-art proof automation of these backend provers;
however, the discrepancies between the polymorphic higher-order logic of Isabelle/HOL
and the monomorphic first-order logic of the backend provers
severely impair Sledgehammer's performance when it comes to inductive theorem proving (ITP).

This is unfortunate for two reasons.
Firstly, many Isabelle users chose Isabelle/HOL
precisely because its higher-order logic is 
expressive enough to specify mathematical objects and procedures involving recursion
without introducing new axioms.
Secondly, 
induction lies at the heart of mathematics and computer science.
For instance, induction is often necessary for reasoning about 
natural numbers, recursive data-structures, such as lists and trees,
computer programs containing recursion and iteration \cite{alan1}.

This is why
ITP remains as a long-standing challenge in computer science, and
its automation is much needed.
Facing the limited automation in ITP,
Gramlich surveyed the problems in ITP and
presented the following prediction in 2005 \cite{gramlich}:

\begin{displayquote}
in the near future, ITP will only be successful for very specialized domains 
for very restricted classes of conjectures.
ITP will continue to be a very challenging engineering process.
\end{displayquote}

\noindent
We address this conundrum with our domain-specific language, \lifter{}.
\lifter{} allows experienced Isabelle users to encode their induction heuristics
in a style independent of problem domains.
\lifter{}'s interpreter mechanically checks 
if a given application of induction is compatible with the induction heuristics
written by experienced users.
Our research hypothesis is that:

\begin{displayquote}
\textit{
it is possible to encode valuable induction heuristics for Isabelle/HOL in \lifter{}
and these heuristics can be valid across diverse problem domains,
because \lifter{} allows for meta-reasoning on applications of induction methods,
without relying on concrete proof goals, their underlying proof states,
nor concrete applications of induction methods.}
\end{displayquote}
\noindent
We developed \lifter{} as an Isabelle theory and
integrated \lifter{} into Isabelle's proof language, Isabelle/Isar,
and its proof editor, Isabelle/jEdit.
This allows for an easy installation process:
to use \lifter{}, users 
only have to import the relevant theory files into their theory files,
using Isabelle's \verb|import| keyword.
Our working prototype is available at GitHub \cite{GitHub}.

The important difference of \lifter{} from other tactic languages, such as Eisbach \cite{eisbach} and Ltac \cite{ltac}, is that
\lifter{} itself is not a tactic language 
but a language to write how one should use Isabelle's existing proof method for induction.
To the best of our knowledge,
\lifter{} is the first language in which
one can write how to use a tactic 
by mechanically analyzing the structures of proof goals
in a style independent of any problem domain.


\section{Induction in Isabelle/HOL} \label{s:background}
To handle inductive problems, modern proof assistants 
offer tools to apply induction. 
For example, Isabelle comes with the \verb|induct| proof method
and the \verb|induction| method
\footnote{Proof methods are the Isar syntactic layer of LCF-style tactics.}.
Nipkow \etal{} proved our ongoing example as follows \cite{concrete_semantics}:

\begin{verbatim}
lemma model_prf:"itrev xs ys = rev xs @ ys"
  apply(induct xs arbitrary: ys) by auto
\end{verbatim}

\noindent
Namely, they applied structural induction on \verb|xs| 
while generalizing \verb|ys| before applying induction 
by passing the string \verb|ys| to the \verb|arbitrary| field.
The resulting sub-goals are:
\begin{verbatim}
1. !!ys. itrev [] ys = rev [] @ ys
2. !!a xs ys. (!!ys. itrev xs ys = rev xs @ ys) ==> 
              itrev (a # xs) ys = rev (a # xs) @ ys
\end{verbatim}

\noindent
where \verb|!!| is the universal quantifier 
and \verb|==>| is the implication in Isabelle's meta-logic.
Due to the generalization, 
the \verb|ys| in the induction hypothesis is quantified within the hypothesis,
and it is differentiated from the \verb|ys| that appears in the conclusion.
Had Nipkow \etal{} omitted \texttt{arbitrary: ys}, 
the first sub-goal would be the same, but the second sub-goal would have been:

\begin{verbatim}
 2. !!a xs. itrev xs ys = rev xs @ ys ==> 
            itrev (a # xs) ys = rev (a # xs) @ ys
\end{verbatim}

\noindent
Since the same \verb|ys| is shared by the induction hypothesis and the conclusion,
the subsequent application of \verb|auto| fails to discharge this sub-goal.

It is worth noting that in general there are multiple equivalently appropriate
combinations of arguments to prove a given inductive problem.
For instance, the following proof snippet shows an alternative
proof script for our example:

\begin{verbatim}
lemma alt_prf:"itrev xs ys = rev xs @ ys"
  apply(induct xs ys rule:itrev.induct) by auto
\end{verbatim}

\noindent
Here we passed the \texttt{itrev.induct} rule to the \verb|rule| field
of the \verb|induct| method and proved the lemma 
by recursion induction \footnote{Recursion induction is also known as functional induction or computation induction.} over \verb|itrev|.
This rule was derived by Isabelle automatically 
when we defined \texttt{itrev},
and it states the following:

\begin{verbatim}
(!!ys. P [] ys) ==>
(!!x xs ys. P xs (x # ys) ==> P (x # xs) ys) ==> 
P a0 a1
\end{verbatim}

\noindent
Essentially, this rule states that
to prove a property \verb|P| of \verb|a0| and \verb|a1|
we have to prove it for two cases where \verb|a0| is the empty list
and the list with at least two elements.
When the \verb|induct| method takes this rule and 
\verb|xs| and \verb|ys| as induction variables,
Isabelle produces
the following sub-goals:

\begin{verbatim}
 1. !!ys. itrev [] ys = rev [] @ ys
 2. !!x xs ys. itrev xs (x # ys) = rev xs @ x # ys ==> 
               itrev (x # xs) ys = rev (x # xs) @ ys
\end{verbatim}

\noindent
where the two sub-goals correspond to the two clauses in the definition 
of \verb|itrev|.

There are other lesser-known techniques to handle difficult inductive problems using the \verb|induct| method, 
and sometimes users have to develop useful auxiliary lemmas manually;
however, for most cases
the problem of how to apply induction 
boils down to the the following three questions:

\begin{itemize}
    \item On which terms do we apply induction?
    \item Which variables do we generalize?
    \item Which rule do we use for recursion induction?
\end{itemize}

\noindent
Isabelle experts resort to induction heuristics to answer such questions 
and decide what arguments to pass to the \verb|induct| method;
however, such reasoning still requires human engineers to carefully investigate the inductive problem at hand.
Moreover, Isabelle experts' induction heuristics are sparsely documented across various documents,
and there was no way to encode their heuristics as programs.
For the wide spread adoption of complete formal verification,
we need a program language to encode such heuristics
and the system to check 
if an invocation of the \verb|induct| method written by an Isabelle novice complies with such heuristics.
We developed \lifter{}, 
taking these three groups of questions as a design space.

\section{Overview and Syntax of \lifter{}} \label{s:overview}

We designed \lifter{} to encode induction heuristics as assertions on invocations of the \verb|induct| method in Isabelle/HOL.
An assertion written in \lifter{} takes the pair of 
a proof goal with its underlying proof state 
and arguments passed to the \verb|induct| method.
When one applies a \lifter{} assertion to an invocation of the \verb|induct| method,
\lifter{}'s interpreter returns a boolean value as the result of the assertion applied to the
proof goals and their underlying proof state.

The goal of a \lifter{} programmer is to write
assertions that implement reliable heuristics.
A heuristic encoded as a \lifter{} assertion is reliable 
when it satisfies the following two properties:
\begin{enumerate}
    \item The \lifter{} interpreter is likely to evaluate 
the assertion to \verb|True|
when the arguments of the \verb|induct| method are
appropriate for the given proof goal.
    \item The \lifter{} interpreter is likely to evaluate the assertion to \verb|False|
when the arguments are inappropriate for the goal.
\end{enumerate}

\begin{figure*}[t]
      \centerline{\includegraphics[width=1.0\linewidth]{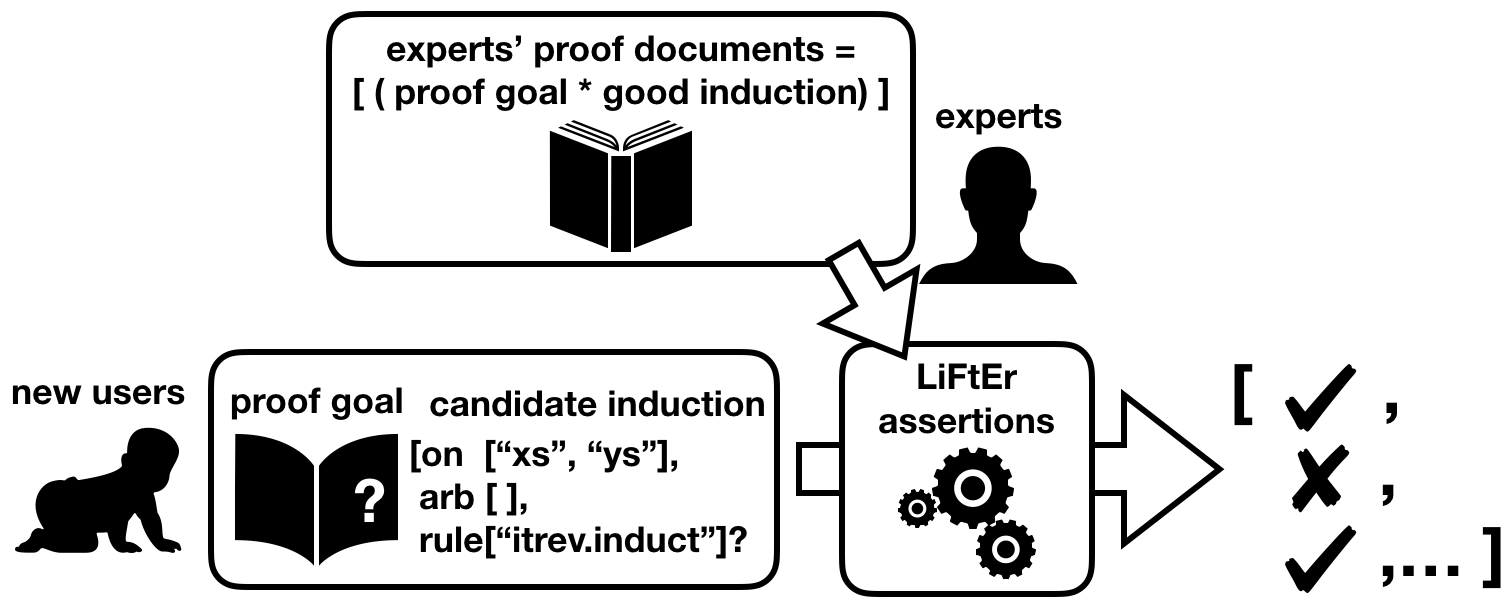}}
      \caption{The Workflow of \lifter{}.}
      \label{fig:system}
\end{figure*}

Fig. \ref{fig:system} illustrates the workflow of \lifter{}.
Firstly, Isabelle experts encode the gist of promising applications of
induction based on experts' proofs.
Note that the heuristics encoded in \lifter{} become applicable to problem domains that
the experts users have not even encountered at the time of writing the assertions. 

When new Isabelle users are facing an inductive problem and are unsure 
if their application of induction is a valid approach or not,
they can apply \lifter{} assertions written by experts using the
\texttt{assert\_LiFtEr} keyword to
their proof goal and 
their candidate arguments. 

\lifter{}'s interpreter checks if the pair of new users' proof goal and candidate
arguments to the \verb|induct| method is compatible with the experts' heuristics.
If the interpreter evaluates the pair to \verb|True|,
Isabelle prints ``\texttt{Assertion succeeded.}'' 
in the Output panel of Isabelle/jEdit \cite{jedit}.
If the interpreter evaluates the pair to \verb|False|,
Isabelle highlights the \texttt{assert\_LiFtEr} in red
and prints ``\texttt{Assertion failed.}'' in the Output panel.

\begin{program*}[!ht]
\verb||\\
\textit{assertion} \verb|:=| \textit{atomic} \texttt{|} \textit{connective} \texttt{|} \textit{quantifier} \texttt{| (} \textit{assertion} \verb|)|\\
\textit{type} \texttt{:=} \texttt{term} \texttt{|} \texttt{term\_occurrence} \texttt{|} \texttt{rule} \texttt{|} \texttt{number}\\
\textit{modifier\_term} \texttt{:= induction\_term | arbitrary\_term}\\
\textit{quantifier} \verb|:=| $\exists$$x$ \verb|:| \textit{type}\verb|.| \textit{assertion}\\ 
\verb|          |\texttt{|} $\forall$$x$ \verb|:| \textit{type}\verb|.| \textit{assertion}\\
\verb|          |\texttt{|} $\exists x$ \verb|:| \verb|term| $\in$ \textit{modifier\_term} \verb|.| \textit{assertion}\\
\verb|          |\texttt{|} $\forall x$ \verb|:| \verb|term| $\in$  \textit{modifier\_term} \verb|.| \textit{assertion}\\
\verb|          |\texttt{|} $\exists$$x$ \verb|:| \verb|term_occurrence| $\in$ $y$ \verb|: term .| \textit{assertion}\\ 
\verb|          |\texttt{|} $\forall x$ \verb|:| \verb|term_occurrence| $\in$ $y$ \verb|: term .| \textit{assertion}\\
\textit{connective} \texttt{:= True | False |}
  \textit{assertion $\lor$ assertion}
\texttt{|} \textit{assertion $\land$ assertion}\\
\verb|            |\textit{assertion $\rightarrow$ assertion} \texttt{|} $\neg$ \textit{assertion}\\
\textit{pattern} \texttt{:=  all\_only\_var | all\_constructor | mixed}\\
\textit{atomic} \verb|:=|\\
\verb|   | \textit{rule} \verb|is_rule_of| \textit{term\_occurrence}\\
\verb|  |\texttt{|} \textit{term\_occurrence} \verb|term_occurrence_is_of_term| \textit{term}\\
\verb|  |\texttt{|} \verb|are_same_term (| \textit{term\_occurrence} \verb|,| \textit{term\_occurrence} \verb|)|\\
\verb|  |\texttt{|} \textit{term\_occurrence}  \texttt{is\_in\_term\_occurrence}  \textit{term\_occurrence}\\
\verb|  |\texttt{|} \verb|is_atomic| \textit{term\_occurrence}\\
\verb|  |\texttt{|} \verb|is_constant| \textit{term\_occurrence}\\
\verb|  |\texttt{|} \verb|is_recursive_constant| \textit{term\_occurrence}\\
\verb|  |\texttt{|} \verb|is_variable| \textit{term\_occurrence}\\
\verb|  |\texttt{|} \verb|is_free_variable| \textit{term\_occurrence}\\
\verb|  |\texttt{|} \verb|is_bound_variable| \textit{term\_occurrence}\\
\verb|  |\texttt{|} \verb|is_lambda| \textit{term\_occurrence}\\
\verb|  |\texttt{|} \verb|is_application| \textit{term\_occurrence}\\
\verb|  |\texttt{|} \textit{term\_occurrence} \verb|is_an_argument_of| \textit{term\_occurrence}\\
\verb|  |\texttt{|} \textit{term\_occurrence} \verb|is_nth_argument_of| \textit{term\_occurrence}\\
\verb|  |\texttt{|} \textit{term} \verb|is_nth_induction_term| \textit{number}\\
\verb|  |\texttt{|} \textit{term} \verb|is_nth_arbitrary_term| \textit{number}\\
\verb|  |\texttt{|} \verb|pattern_is (| \textit{number} \texttt{,} \textit{term\_occurrence} \texttt{,} \textit{pattern} \verb|)|\\
\verb|  |\texttt{|} \verb|is_at_deepest| \textit{term\_occurrence}\\
\verb|  |\texttt{|} \texttt{\dots}\\
\caption{The Abstract Syntax of \lifter{}.}
\label{p:sytax}
\end{program*}

Program \ref{p:sytax} shows the essential part of \lifter{}'s abstract syntax.
\lifter{} has four types of variables:
\numbvar{}, \rulevar{}, \trmvar{}, and \occvar{}.
A value of type \numbvar{} is a natural number from $0$ to the maximum of the following two numbers:
the number of terms appearing in the proof goals at hand, and
the maximum arity of constants appearing in the proof goals.
A value of type \rulevar{} corresponds to a name of an auxiliary lemma passed
to the \verb|induct| method as an argument in the \verb|rule| field.

The difference between \trmvar{} and \occvar{} is crucial:
a value of \trmvar{} is a term appearing in proof goals,
whereas a value of \occvar{} is an \textit{occurrence} of such terms.
It is important to distinguish terms and term occurrences
because the \verb|induct| method in Isabelle/HOL only allows its users to specify induction terms
but it does not allow us to specify on which occurrences of such terms we intend to apply induction.

The connectives, $\land$, $\lor$, $\neg$, and $\rightarrow$ correspond to
conjunction, disjunction, negation, and implication in the classical logic, respectively;
and $\rightarrow$ admits the principle of explosion.

\lifter{} has four essential quantifiers and two quantifiers as syntactic sugar.
As is often the case, $\forall$ quantifies over variables universally,
and $\exists$ stands for the existence of a variable it binds.
Again, it is important to notice the difference between the quantifiers over \trmvar{} and
the ones over \occvar{}.
For example,
$\forall{} \_$\verb|.| $\in$ \trmvar{} quantifies all sub-terms appearing in the proof goals, whereas
$\forall{} \_$\verb|.| $\in$ \occvar{} quantifies all \textit{occurrences} of such sub-terms.
Quantified variables restricted to \texttt{induction\_term} 
by the membership function $\in$ are quantified over all terms passed
to the \texttt{induct} method as induction terms, while
quantified variables restricted to \texttt{arbitrary\_term}
are quantified over all terms passed to the \texttt{induct} method
as arguments in the \texttt{arbitrary} field. 

Some atomic assertions judge properties of term occurrences,
and some judge the syntactic structure of proof goals 
with respect to certain terms, their occurrences, or certain numbers.
While most atomic assertions work on the syntactic structures
of proof goals,
\verb|Pattern| provides a means to describe a limited amount of semantic information of proof goals
since it checks how terms are defined.
Section \ref{s:example} explains the meaning of important atomic assertions through \lifter{}'s standard heuristics.

Attentive readers may have noticed that 
\lifter{}'s syntax does not cover any user-defined types or constants. 
This absence of specific types and constants is our intentional choice 
to promote induction heuristics that are valid across various problem domains:
it encourages \lifter{} users to
write heuristics that are 
not specific to particular data-types or functions.
And \lifter{}'s interpreter can check if an application of the \verb|induct| method
is compatible with a given \lifter{} heuristic 
even if the proof goal involves user-defined data-types and functions
even though such types and functions are unknown to 
the \lifter{} developer or the author of the heuristic
but come into existence in the future 
only after developing \lifter{} and such heuristic.

\section{\lifter{}'s Standard Heuristics}\label{s:example}
This section presents \lifter{}'s standard heuristics and
illustrates how to use those atomic assertions and quantifiers to encode induction heuristics.

\subsection{Heuristic 1: Induction terms should not be constants.} \label{s:example1}
Let us revise the first example lemma
about the equivalence of two reverse functions,
\verb|itrev| and \verb|rev|.
One naive induction heuristic would be 
\textit{``any induction term should not be a constant''}
\footnote{This \textit{naive heuristic} is not very reliable:
there are cases where the \texttt{induct} method takes terms involving constants
and apply induction appropriately 
by automatically introducing induction variables.
See Concrete Semantics \cite{concrete_semantics} for more details.}
In \lifter{}, we can encode this heuristic as the following assertion \footnote{For better readability we omit parentheses
where the binding of terms is obvious from indentation.}:

\begin{alltt}
\( \forall t1 \): term \(\in\) induction_term.
  \( \exists to1 \): term_occurrence.
       ( \(to1 \) term_occurrence_is_of_term \(t1\) )
     \(\land\)
       \(\neg\) ( is_constant \(to1\) )
\end{alltt}

\noindent
Note the use of quantifiers over \texttt{induction\_term}s and \occvar{}s:
when \lifter{} handles induction terms,
\lifter{} treats them as terms,
but it is often necessary to analyze the \textit{occurrences} of these terms in the proof goal
to decide how to apply induction.
In our example lemma, 
\verb|xs| is a variable, which appears twice:
once as the first argument of \verb|itrev|,
and once as the first argument of \verb|rev|.
With this in mind, the above assertion reads as follows:

\begin{displayquote}
for all induction terms, named $t1$,
 there exists a term occurrence, named $to1$, such that
   $to1$ is an occurrence of $t1$
  and
   $to1$ is not a constant.
\end{displayquote}

\noindent
Now we compare this heuristic 
with the model proof by Nipkow \etal{}

The only induction term, \texttt{xs}, has two occurrences in the proof goal
both as variables.
Therefore, if we apply this \lifter{} assertion to the model proof,
\lifter{}'s interpreter acknowledges that
the model proof complies with the induction heuristic defined above.

\begin{figure*}[t]
      \centerline{\includegraphics[width=1.0\linewidth]{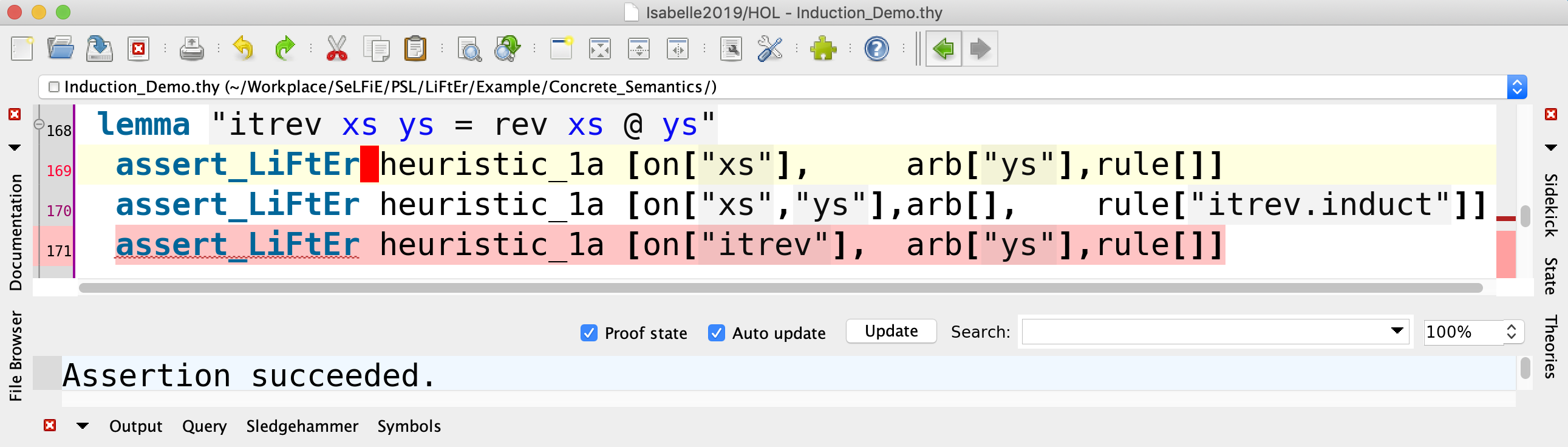}}
      \caption{The User Interface of \lifter{}.}
      \label{fig:screenshot}
\end{figure*}

Fig. \ref{fig:screenshot} shows the user interface of \lifter{}.
In the second line where the cursor is staying,
\lifter{}'s interpreter executes the aforementioned reasoning and concludes that
the model proof by Nipkow \etal{} is compatible with this heuristic,
printing ``\texttt{Assertion succeeded.}'' in the Output panel.
On the contrary, the fourth line applies the same heuristic to 
another possible combination of arguments to the \verb|induct| method
(\texttt{induct itrev arbitrary: ys}) and
concludes that this candidate induction is not compatible with our heuristic
because \texttt{itrev} is a constant.
\lifter{} also highlights this line in red to warn the user.

It is a common practice to analyze occurrences of specific terms
when describing induction heuristics.
Therefore, we introduced two pieces of syntactic sugar to avoid boilerplate code:
$\exists{} \_$ \verb|:| \occvar{} $\in \_$ \verb|: term| and
$\forall{} \_$ \verb|:| \occvar{} $\in \_$ \verb|: term|.
Both of these quantify over term occurrences of a particular term
rather than all term occurrences in the proof goal at hand.
Using $\exists{} \_$ \verb|:| \occvar{} $\in \_$ \verb|: term|, we can shrink the above assertion from
5 lines to 3 lines as follows:
\begin{alltt}
\(\forall t1 \): induction_term. 
  \(\exists to1 \): term_occurrence \(\in t1 \): term.
    \( \neg \) ( is_constant \( to1 \))
\end{alltt}

\noindent
In English, this reads as follows:
\begin{displayquote}
for all induction terms, named $t1$,
 there exists an occurrence of $t1$, named $to1$, such that
  $to1$ is not a constant.
\end{displayquote}

\subsection{Heuristic 2. Induction terms should appear at the bottom of syntax trees.}\label{s:bottom}
Not applying induction on a constant would sound a plausible heuristic,
but such heuristic is not very useful.

In this sub-section, we encode an induction heuristic that analyzes
not only the properties of the induction terms
but also the location of their occurrences 
within the proof goal at hand.
When attacking inductive problems with many variables,
it is sometimes a good attempt 
to apply induction on variables that appear at the bottom of the syntax tree
representing the proof goal.
We encode such heuristic using \texttt{is\_at\_deepest}
as the following \lifter{} assertion:

\begin{alltt}
\(\forall t1 \): induction_term. 
  \(\exists to1 \): term_occurrence \(\in t1 \): term.
      is_atomic \(to1\) \( \rightarrow \) is_at_deepest \(to1\)
\end{alltt}

\noindent
In English, this assertion reads as follows:
\begin{displayquote}
for all induction terms, named $t1$,
 there exists an occurrence of $t1$, named $to1$, such that
  if
   $to1$ is an atomic term
  then
   $to1$ lies at the deepest layer in the syntax tree that represents the proof goal.
\end{displayquote}

\noindent
We used the infix operator, $\rightarrow$, to add the condition that
we consider only the induction terms that are atomic terms.
An atomic term is either a constant, free variable, schematic variable, or variable bound by a lambda abstraction.
We added this condition because
it makes little sense to check if the induction term resides at the bottom of the syntax tree
when an induction term is a compound term:
such compound terms have sub-terms at lower layers.

\lifter{}'s interpreter acknowledges that
the model solution provided by Nipkow \etal{} complies with this heuristic
when applied to this lemma:
there is only one induction term, \verb|xs|,
and \texttt{xs} appears as an argument of \texttt{rev} 
on the right-hand side of the equation in the lemma at the lowest layer of this syntax tree.

\subsection{Heuristic 3. All induction terms should be arguments of the same occurrence of a recursively defined function.}

Probably, it is more meaningful to analyze where induction terms reside in the proof goal
with respect to other terms in the goal.
More specifically, one heuristic for promising application of induction would be
\textit{``apply induction on terms that appear as arguments of 
the same occurrence of a recursively defined function''}.
We encode this heuristic using \lifter{}'s atomic assertions, 
\texttt{is\_recursive\_constant} and \texttt{is\_an\_argument\_of}, as follows:

\begin{alltt}
\(\exists t1 \): term. 
  \(\exists to1 \): term_occurrence \(\in t1\) : term.
    \(\forall  t2\): term \(\in\) induction_term.
      \(\exists to2 \): term_occurrence \(\in t2\) : term.
          is_recursive_constant \(to1\) \( \land \) \(to2\) is_an_argument_of \(to1 \)
\end{alltt}

\noindent
where \texttt{is\_recursive\_constant} checks
if a constant is defined recursively or not,
and \texttt{is\_an\_argument\_of} takes two term occurrences and checks if
the first one is an argument of the second one.

Note that using \texttt{is\_recursive\_constant}
this assertion checks not only the syntactic information of the proof goal at hand,
but it also extracts an essential part of the semantic information of constants appearing in the goal,
by investigating how these constants are defined in the underlying proof context.
As a whole, this assertion reads as follows:
\begin{displayquote}
there exists a term, named $t1$, such that
 there exists an occurrence of $t1$, named $to1$, such that
  for all induction terms, named $t2$,
   there exists an occurrence of $t2$, named $to2$, such that
    $to1$ is defined recursively
    and
    $to2$ appears as an argument of $to1$.
\end{displayquote}
Attentive readers may have noticed that 
we quantified over induction terms within 
the quantification over $to1$, 
so that this induction heuristic checks 
if all induction terms occur as arguments of the same constant.

The \lifter{} interpreter confirms that the model proof is compatible with
this heuristic as well:
the constant, \texttt{itrev}, is defined recursively
and has an occurrence that takes 
the only induction variable \texttt{xs} as the first argument.

\subsection{Heuristic 4. 
One should apply induction on the nth argument of a function
where the nth parameter in the definition of the function
always involves a data-constructor.}

The previous heuristic checks if 
all induction terms are arguments of the same occurrence of a 
recursively defined function.
Sometimes we can even estimate on which arguments of such function
we should apply induction
by inspecting the definition of the function more carefully.

We introduce two 
constructs to support this style of reasoning:
\texttt{is\_nth\_argument\_of} 
and \texttt{pattern\_is}.
\texttt{is\_nth\_argument\_of} takes a term occurrence, a number, and another term occurrence,
and it checks if the first term occurrence is the $n$th argument of the second term occurrence
where counting starts at $0$. 
\texttt{pattern\_is} takes a number, a term occurrence, one of three \textit{pattern}s:
\texttt{all\_only\_var}, \texttt{all\_constructor}, and \texttt{mixed}. 
Each of such patterns describes how the term is defined.

For example, \texttt{pattern\_is (}$n$\texttt{,} $to$\texttt{, all\_only\_var)} denotes that
the $n$th parameter is always a variable on the left-hand side of
the definition of the term that has the term occurrence, $to$.
Likewise, \texttt{all\_constructor} denotes the case where the corresponding parameter of the definition
of a particular constant always involves a data-constructor, whereas
\texttt{mixed} denotes that the corresponding parameter is a variable in some clauses but involves
a data-constructor in other clauses.
With these atomic assertions in mind, 
we write the following \lifter{} assertion:

\begin{alltt}
  \(\neg \)(\(\exists r1\) : rule. True)
\(\rightarrow \)
  \(\exists t1\) : term.
    \(\exists to1\) : term_occurrence \(\in{}t1\) : term.
        is_recursive_constant \(to1\)
      \(\land\)
        \(\forall t2\) : term \(\in \) induction_term.
          \(\exists to2\) : term_occurrence \(\in{}t2\) : term.
            \(\exists n \): number.
                pattern_is (\(n\), \(to1\), all_constructor)
              \(\land\)
                is_nth_argument_of (\(to2\), \(n\), \(to1\))
\end{alltt}

\noindent
This roughly translates to the following English sentence:

\begin{displayquote}
if there is no argument in the \texttt{rule} field in the \texttt{induct} method,
then
 there exists a recursively defined constant, $t1$, 
 with an occurrence, $to1$, such that
  for all induction terms $t2$,
   there exists an occurrence, $to2$, of $t2$, such that
    there exists a number, $n$, such that
     the $n$th parameter involves a data-constructor in all the clauses
     of the definition of $t1$, and
    $to2$ appears as the $n$th argument of $to1$ in the proof goal.
\end{displayquote}

\noindent
Note that we added 
$\neg$ \verb|(|$\exists r1$ \texttt{: rule. True)}
to focus on the case
where the \verb|induct| method does not take any auxiliary lemma in the \verb|rule| field
since this heuristic is known to be less reliable if there is an auxiliary lemma
passed to the \verb|induct| method.

\lifter{}'s interpreter confirms that Nipkow's proof about \verb|itrev| and \verb|rev|
conforms to this heuristic:
there exists an occurrence of \verb|itrev|, such that
    \verb|itrev| is recursively defined
  and
    for the only induction term, \verb|xs|, 
      there is an occurrence of \verb|xs| on the left-hand side of the proof goal, such that
        \verb|itrev|'s first parameter involves data-constructor in all clauses of its definition, and
        this occurrence of \verb|xs| appears as the first argument of the occurrence of \verb|itrev| in the goal
        \footnote{Note that in reality the counting starts at $0$ internally. 
        Therefore, ``the first argument'' in this English sentence is processed as the $0$th argument within \lifter{}.}.

\subsection{Heuristic 5. Induction terms should appear as arguments of a function 
that has a related \texttt{.induct} rule in the \texttt{rule} field.}
When the \texttt{induct} method takes an auxiliary lemma in the \texttt{rule} field
that Isabelle automatically derives from the definition of a constant,
it is often true that we should apply induction on terms that appear as arguments of
an occurrence of such constant.

See, for example, our alternative proof, \verb|alt_prf|, for our ongoing example theorem.
When Nipkow \etal{} defined the \texttt{itrev} function with the \texttt{fun} keyword,
Isabelle automatically derived the auxiliary lemma \texttt{itrev.induct},
and the occurrence of \texttt{itrev} on the left-hand side of the equation 
takes \texttt{xs} and \texttt{ys} as its arguments.
Furthermore, the alternative proof passes \texttt{xs} and \texttt{ys} to the \verb|rule| field in the 
same order they appear as the arguments of the occurrence of \texttt{itrev} in the proof goal.

We introduce \texttt{is\_rule\_of} to relate a term occurrence with an auxiliary lemma
passed to the \verb|rule| field.
\texttt{is\_rule\_of} takes a term occurrence and an auxiliary lemma
in the \texttt{rule} field of the \verb|induct| method,
and it checks if
the rule was derived by Isabelle at the time of defining the term.
Moreover, we introduce \texttt{is\_nth\_induction\_term}, 
which allows us to specify the order of induction terms
passed to the \verb|induct| method:
\texttt{is\_nth\_induction\_term} takes a term and a number, 
and it checks if
the term is passed to the \verb|induct| method as the $n$th induction term.
Using these constructs, we can encode the aforementioned heuristic as follows:

\begin{alltt}
  \(\exists r1\) : rule. True
\(\rightarrow \)
  \(\exists r1\) : rule.
    \(\exists t1\) : term.
      \(\exists to1\) : term_occurrence \(\in t1\) : term.
          \(r1\) is_rule_of \(to1\)
        \(\land\)
          \(\forall t2\) : term \(\in\) induction_term.
            \(\exists to2\) : term_occurrence \(\in t2\) : term.
              \(\exists n\) : number.
                  is_nth_argument_of (\(to2\), \(n\), \(to1\))
                \(\land\)
                  \(t2\) is_nth_induction_term \(n\)
\end{alltt}

\noindent
As a whole this \lifter{} assertion checks if the following holds:

\begin{displayquote}
if there exists a rule, $r1$, 
in the \texttt{rule} field of the \verb|induct| method,
then there exists a term $t1$ with an occurrence $to1$, 
such that
$r1$ is derived by Isabelle when defining $t1$, and
for all induction terms $t2$,
there exists an occurrence $to2$ of $t2$ such that,
there exists a number $n$, such that
$to2$ is the $n$th argument of $to1$ and that
$t2$ is the $n$th induction terms passed to the \verb|induct| method.
\end{displayquote}

\noindent
Our alternative proof is compatible with this heuristic:
there is an argument, \texttt{itrev.induct}, in the \verb|rule| field,
and the occurrence of its related term, \verb|itrev|, in the proof goal takes
all the induction terms, \verb|xs| and \verb|ys|, as its arguments in the same order.

\subsection{Heuristic 6. Generalize variables in induction terms.}\label{s:example6}

Isabelle's \verb|induct| method offers the \verb|arbitrary| field, so that
users can specify which terms to be generalized in induction steps;
however, it is known to be a hard problem to decide which terms to generalize.

Of course \lifter{} cannot provide you with a decision procedure to determine
which terms to generalize, but it allows us to describe heuristics
to identify variables that are likely to be generalized by experienced Isabelle users.
For example, experienced users know that
it is usually a bad idea to pass induction terms themselves to the \verb|arbitrary| field.
We also know that
it is often a good idea to generalize variables appearing within induction terms
if induction terms are compound terms.

We can encode the former heuristic using \verb|are_same_term|,
which checks if two terms are the same term or not.
For instance, we can write the following:

\begin{alltt}
\(\forall t1\) : term \(\in\) arbitrary_term. 
  \(\neg\) (\(\exists t2\) : term \(\in\) induction_term. are_same_term (\(t1\), \(t2\)))
\end{alltt}

\noindent
By now, it should be easy to see that this assertion checks if the following holds:

\begin{displayquote}
for all terms in the \verb|arbitrary| field,
 there is no induction term of the same term in the \verb|induct| method.
\end{displayquote}

\noindent
The latter heuristic involves the description of the term structure 
constituting the proof goal.
For this purpose we use \texttt{is\_in\_term\_occurrence}
to check if a term occurrence resides within another term occurrence.
With this construct, we can encode the latter heuristic as follows:

\begin{alltt}
\(\exists t1\) : term \(\in\) induction_term.
  \(\exists to1\) : term_occurrence \(\in t1\) : term.
    \(\forall t2\) : term.
      \(\exists to2\) : term_occurrence \(\in t2\) : term.
          ( \(to2\) is_in_term_occurrence \(to1\) \(\land\) is_free_variable \(to2\) )
        \(\rightarrow \)
          \(\exists t3\) : term \(\in\) arbitrary_term. are_same_term (\(t2\), \(t3\))
\end{alltt}

\noindent
Again, we used the implication (\verb|_| $\rightarrow$ \verb|_|) to avoid applying this generalization heuristics to the cases
without compound induction terms.

\subsection{Apply all assertions
using the \texttt{test\_all\_LiFtErs} command.}
\begin{figure*}[t]
      \centerline{\includegraphics[width=1.0\linewidth]{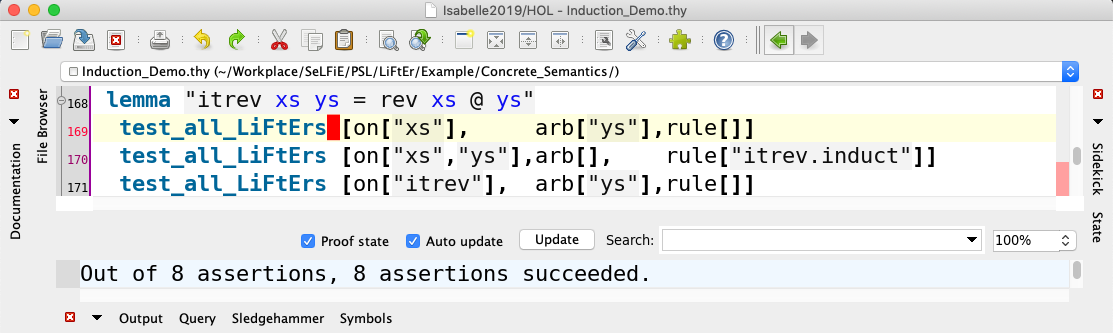}}
      \caption{The \texttt{test\_all\_LiFtErs} command.}
      \label{fig:test_all_lifters}
\end{figure*}

In this section we have written eight assertions
(two assertions from each of Section \ref{s:example1}
and Section \ref{s:example6}).
To exploit all the available \lifter{} assertions,
we developed the \texttt{test\_all\_LiFtErs} command.
The \texttt{test\_all\_LiFtErs} command
first takes a combination of induction arguments to the \verb|induct| method.
Then, it applies all the available \lifter{} assertions
to the pair of the combination of arguments 
and the proof goal at hand.
Finally, it counts how many assertions return \verb|True|.
For example, 
the second line in Fig. \ref{fig:test_all_lifters}
executed the eight available assertions
to the combination of arguments 
(\texttt{[on["xs"], arb["ys"], rule[]]}) and the proof goal.
The output panel shows the result:
\texttt{Out of 8 assertions, 8 assertions succeeded.}
This indicates that
the model proof by Nipkow is indeed a good solution
in terms of all the heuristics we discussed in this section.

\section{Induction Heuristics Across Problem Domains}\label{s:cross_domain}
In Section \ref{s:example} we wrote eight assertions in \lifter{}.
When writing these eight assertions, we emphasized that 
none of them is specific to the data structure \verb|list| or
the function \verb|itrev| appearing in the proof goal.
In this section we demonstrate that the \lifter{} assertions written in Section \ref{s:example}
are applicable across domains, 
taking an inductive problem from a completely different domain as an example.
The following code is the formalization of a simple stack machine from Concrete Semantics \cite{concrete_semantics}:

\begin{verbatim}
type_synonym vname = string
type_synonym val   = int
type_synonym state = "vname => val"
datatype instr     = LOADI val | LOAD vname | ADD
type_synonym stack = "val list"

fun exec1 :: "instr => state => stack => stack" where
  "exec1 (LOADI n) _ stk        =  n       # stk"
| "exec1 (LOAD x)  s stk        =  s(x)    # stk"
| "exec1  ADD      _ (j#i#stk)  =  (i + j) # stk"

fun exec :: "instr list => state => stack => stack" where
  "exec []     _ stk = stk"
| "exec (i#is) s stk = exec is s (exec1 i s stk)"
\end{verbatim}

\noindent
\verb|exec1| defines how the stack machine in a certain state
transforms a given stack into a new one
by executing one instruction, whereas
\verb|exec| specifies how the machine executes a series of instructions one by one.
Nipkow \etal{} proved the following lemma using structural induction.

\begin{verbatim}
lemma exec_append_model_prf[simp]:
  "exec (is1 @ is2) s stk = exec is2 s (exec is1 s stk)"
  apply(induct is1 arbitrary: stk) by auto
\end{verbatim}

\noindent
This lemma states that
executing a concatenation of two lists of instructions 
in a state to a stack produces the same stack
as executing the first list of the instructions first in the same state 
to the same stack and
executing the second list again in the same state again 
but to the resulting new stack.
As in the case with the equivalence of two reverse functions, 
there is also an alternative proof based on recursion induction:

 \begin{verbatim}
lemma exec_append_alt_prf:
  "exec (is1 @ is2) s stk = exec is2 s (exec is1 s stk)"
  apply(induct is1 s stk rule:exec.induct) by auto
\end{verbatim}

\noindent
where \texttt{exec.induct} is automatically derived by Isabelle 
when defining \texttt{exec}.
Now we check if 
the heuristics from Section \ref{s:example} correctly recommend these proofs.

\paragraph{Heuristic 1.}
Both \execMdl{} and \execAlt{} comply with this heuristic. 
For example, \verb|is1| is the only induction term in \execMdl{}, 
and it has occurrences in the proof goal, 
where it occurs as a variable.

\paragraph{Heuristic 2.}
\execMdl{} complies with the second heuristic:
its only induction term, \verb|is1|, occurs 
at the bottom of the syntax tree as a variable,
which is an atomic term.
\execAlt{} also complies with this heuristic:
\verb|is1|, \verb|s|, and \verb|stk| as
the arguments of the inner \verb|exec| 
on the right-hand side of the equation
are all atomic terms at the deepest layer
of the syntax tree.

\paragraph{Heuristic 3.}
Both proof scripts comply with this heuristic.
For example, 
the inner occurrence of \verb|exec| on the right-hand side
of the equation takes
all the induction terms of the alternative proof
(namely, \verb|is1|, \verb|s|, and \verb|stk|)
as its arguments.

\paragraph{Heuristic 4.}
This heuristic works for both proof scripts,
but it explains the model answer particularly well:
it has a recursively defined constant, \verb|exec|,
and the inner occurrence of \verb|exec| on the right-hand 
side of the equation has an occurrence that takes the only
induction term \verb|is1| as its first argument,
and the first parameter of \verb|exec| always involve 
a data-constructor in the definition of \verb|exec|.

\paragraph{Heuristic 5.}
This heuristic also works for both proof scripts,
but it fits particularly well with the alternative answer:
the rule \texttt{exec.induct} is derived by Isabelle 
when defining \verb|exec|,
while \verb|exec| has an occurrence as 
part of the third argument of another \verb|exec|
on the right-hand side of the equation,
and this inner occurrence of \verb|exec| takes 
all the induction terms (\verb|is1|, \verb|s|, and \verb|stk|)
in the same order.

\paragraph{Heuristic 6.}
None of our proofs involve induction on a compound term,
making the second assertion in Section \ref{s:example6} rather irrelevant,
whereas the first assertion in Section \ref{s:example6} explains the model answer well:
the only generalized term, \verb|stk|, does not appear as an induction term.

\section{Real World Example}\label{s:difficult}
\begin{program*}[!ht]
\begin{alltt}

datatype com = 
  SKIP                            (*No-op*)
(*Assignment*)
| AssignIdx vname aexp aexp       (*Assign to index in array*)
| ArrayCpy vname vname            (*Copy whole array*)
| ArrayClear vname                (*Clear array*)
| Assign_Locals "vname => val"    (*Assign all local variables*)
(*Block*)
| Seq    com  com                 (*Sequential composition*)
| \(\dots\)
\end{alltt}

\begin{alltt}
fun small_step :: "program => com \(\times\) state => (com \(\times\) state) option" where
  "small_step \(\pi\) ((AssignIdx x i a, s) = 
     Some (SKIP, s(x := (s x)(aval i s := aval a s)))"
| "small_step \(\pi\) (ArrayCpy x y, s)    = Some (SKIP, s(x := s y))"  
| "small_step \(\pi\) (ArrayClear x, s)    = Some (SKIP, s(x := (\(\lambda\)_. 0)))"
| "small_step \(\pi\) (Assign_Locals l, s) = Some (SKIP, <l|s>)"
| "small_step \(\pi\) (SKIP ;; c, s)       = Some (c, s)"
| "small_step \(\pi\) (c1 ;; c2, s)        = (case small_step \(\pi\) (c1, s) of 
     Some (c1', s') => Some (c1' ;; c2, s') | _ => None)"
| \(\dots\)
\end{alltt}
\begin{alltt}
inductive small_steps :: 
  "program => com \(\times\) state => (com \(\times\) state) option => bool" where 
  "small_steps \(\pi\) cs (Some cs)"
| "small_step \(\pi\) cs = None \(\longrightarrow\) small_steps \(\pi\) cs None"
| "small_step \(\pi\) cs = Some cs1 \(\longrightarrow\) 
   small_steps \(\pi\) cs1 cs2 \(\longrightarrow\) small_steps \(\pi\) cs cs2"
\end{alltt}

\begin{alltt}
lemma smalls_seq:
  "small_steps \(\pi\) (c, s) (Some (c', s')) \(\Longrightarrow\) 
   small_steps \(\pi\) (c ;; cx, s) (Some (c';; cx, s'))"
  apply (induct \(\pi\) "(c, s)" "Some (c', s')" 
         arbitrary: c s c' s' rule: small_steps.induct)
  apply (auto dest: small_seq intro: small_steps.intros)
  by (metis option.simps(1) prod.simps(1) 
            small_seq small_step.simps(31) small_steps.intros(3))
\end{alltt}
\caption{A Proof about the Semantics of an Imperative Language, IMP2.}
\label{p:difficult_proof}
\end{program*}

In Section \ref{s:example} and Section \ref{s:cross_domain},
we introduced simple \lifter{} assertions applied to smaller problems.
For example, all induction terms in the examples were variables,
even though Isabelle's \verb|induct| method can induct on non-atomic terms.

Program \ref{p:difficult_proof} is a more challenging proof 
about a formalization of an imperative language, IMP2 \cite{imp2},
from the Archive of Formal Proofs \cite{AFP}.
Due to the space constraints, 
we refrain ourselves from presenting the complete formalization of IMP2
but focus on the essential part of the proof document.

In this project, Lammich \etal{} proved the equivalence between IMP2's
big-step semantics and small-step semantics. 
\texttt{smalls\_seq} in Program \ref{p:difficult_proof} 
is an auxiliary lemma useful to prove the equivalence.
The proof of \texttt{smalls\_seq} appears to be somewhat similar to
that of \texttt{alt\_prf} in Section \ref{s:background} and 
\texttt{exec\_append\_alt\_prf} in Section \ref{s:cross_domain}:
\texttt{smalls\_seq}'s proof uses the auxiliary lemma 
\texttt{small\_steps.induct}, which Isabelle derived automatically
when Lammich \etal{} defined \texttt{small\_steps}.
Furthermore, the three induction terms, \texttt{$\pi$}, \texttt{(c, s)}, and \texttt{Some (c', s')}, 
are the arguments of one occurrence of \texttt{small\_steps}.

The difference from the preceding examples is 
the generalization of four free variables appearing in induction terms:
in Program \ref{p:difficult_proof}, 
\verb|c| and \verb|s| appear within \texttt{(c, s)}, 
while \verb|c'| and \verb|s'| appear within \texttt{Some (c', s')}.
As we discussed in Section \ref{s:example6},
when applying induction on non-atomic terms in Isabelle/HOL
it is often a good idea to 
generalize free variables appearing within such non-atomic induction terms.

To encode such heuristic, 
we strengthened Example 5 in Section \ref{s:example} 
using the \texttt{is\_in\_term\_occurrence} assertion.
Program \ref{p:difficult_assertion} checks
if any induction term is non-atomic and contains a free variable,
all such free variables are generalized in the
\texttt{arbitrary} field.
Note that 
\lifter{}'s interpreter evaluates the universal quantifier
over $to3$ to \verb|True|
when all induction terms are atomic,
since $to3$ \texttt{term\_occurrence\_is\_of\_term} $t3$ is guarded
by \texttt{$\neg$ ( is\_atomic $to2$ )},
making this assertion valid even for 
the cases where induction terms are atomic variables.

\begin{program*}[!ht]
\begin{alltt}

  \(\exists r1\) : rule. True
\(\rightarrow \)
  \(\exists r1\) : rule.
    \(\exists t1\) : term.
      \(\exists to1\) : term\_occurrence \(\in t1\) : term.
          \(r1\) is_rule_of \(to1\)
        \(\land\)
          \(\forall t2\) : term \(\in\) induction_term.
            \(\exists to2\) : term_occurrence \(\in t2\) : term.
                \(\exists n1\) : number.
                    is_nth_argument_of (\(to2\), \(n1\), \(to1\))
                  \(\land\)
                    \(t2\) is_nth_induction_term \(n1\)
              \(\land\)
                \(\forall to3\) : term_occurrence.
                      \(\neg\) ( is_atomic \(to2\) )
                    \(\land\)
                      is_free_variable \(to3\)
                    \(\land\)
                      \(to3\) is_in_term_occurrence \(to2\)
                  \(\rightarrow\)
                    \(\exists{} t3\) : arbitrary_term.
                      \(to3\) term_occurrence_is_of_term \(t3\)
\end{alltt}
\caption{An Assertion for the Generalization of Variables in Induction Terms.}
\label{p:difficult_assertion}
\end{program*}

\section{Conclusions, Related and Future Work}\label{s:related_work}
ITP has been considered as
a very challenging task. 
To address this issue, we presented \lifter{}.
\lifter{} is a domain-specific language in the sense that 
we developed \lifter{} to encode induction heuristics;
however, heuristics written in \lifter{} are often not specific to any problem domains.
To the best of our knowledge, \lifter{} is the first programming language
developed to capture induction heuristics across problem domains,
and its interpreter is the first system that
executes meta-reasoning on interactive inductive theorem proving.


The recent development in proof automation for higher-order logic
takes the meta-tool approach.
Gauthier \etal{}, for example, developed an automated tactic prover,
TacticToe, on top of HOL4 \cite{tactictoe}.
TacticToe leanrs how human engineers used tactics and
applies the knowledge to execute a tactic based Monte Carlo tree search.
To automate proofs in Coq \cite{coq},
Komendantskaya \etal{} developed ML4PG \cite{ml4pg2}. 
ML4PG uses recurrent clustering to mine a proof database and 
attempts to find a tactic-based proof for a given proof goal.
Both of them try to identify useful lemmas or hypotheses as arguments of a tactic;
however, they do not identify promising terms as arguments of a tactic
even though identifying such terms is crucial to apply induction effectively.

The most well-known approach for ITP is 
called the Boyer-Moore waterfall model \cite{waterfall}.
This approach was invented for a first-order logic on Common Lisp.
Most waterfall provers attempt to 
apply six proof techniques (simplification, destructor elimination, 
cross-fertilization, generalization, elimination of irrelevance, and induction) in a fixed order,
store the resulting sub-goals in a pool,
and keep applying these techniques until the pool becomes empty.

ACL2 \cite{acl2} is the most commonly used waterfall model based prover,
which has achieved industrial-scale proofs \cite{acl2_industry}.
When deciding how to apply induction,
ACL2 computes a score, called \textit{hitting ratio}, 
to estimate
how good each induction scheme is for the term which it accounts for
and proceeds with the induction scheme with the highest hitting ratio \cite{acl_book, induction}.

Compared to the hitting ratio used in the waterfall model,
\lifter{}'s atomic assertions let us analyze the structures of proof goals directly
while \lifter{}'s quantifiers let us keep \lifter{} assertions non-specific to any problem.
While ACL2 produces many induction schemes 
and computes their hitting ratios,
\lifter{} assertions do not directly produce induction schemes but analyze the given proof goal
and the arguments passed to the \texttt{induct} method, 
re-using Isabelle's existing tool to (implicitly) produce induction principles.
We consider \lifter{}'s approach to be a reasonable choice,
since it extends the usability of the already well-developed proof assistant, Isabelle/HOL, 
while avoiding to reinvent the mechanism to produce induction principle.

Furthermore, the choice of Isabelle/HOL as the host system of \lifter{}
allowed us to take advantage of human interaction more aggressively
both from Isabelle experts and new Isabelle users:
Isabelle experts can encode their own heuristics 
since \lifter{} is a language,
and new Isabelle users can inspect the results of \lifter{} assertions
and decide how to attack their proof goals
instead of following the fixed order of six proof techniques as in the waterfall model.

Heras \etal{} used ML4PG learning method to find patterns
to generalize and transfer inductive proofs from one domain to another in ACL2
\cite{ml4pg_acl}.
Jiang \etal{} followed the waterfall model
and ran multiple waterfalls \cite{jiang} 
to automate ITP in HOL light \cite{hollight}.
However, when deciding induction variables,
they naively picked the first free variable with recursive type and 
left the selection of appropriate induction variables 
as future work.

To determine induction variables automatically,
we developed a proof strategy language 
\verb|PSL| and its default proof strategy, \verb|try_hard| for Isabelle/HOL \cite{psl}. 
\verb|PSL| tries to identify useful arguments for the \verb|induct| method by conducting a depth-first search.
Sometimes it is not enough to pass arguments to the \verb|induct| method,
but users have to specify necessary auxiliary lemmas before applying induction.
To automate such labor-intensive work,
\pgt{} \cite{pgt}, a new extension to \psl{}, 
produces many lemmas 
by transforming the given proof goal
while trying to identify a useful one in a goal-oriented manner.

The drawback of \psl{} and \pgt{} is that 
they cannot produce recommendations
if they fail to complete a proof search:
when the search space becomes enormous, 
neither \psl{} and \pgt{} gives any advice to Isabelle users.

\pamper{} \cite{pamper}, on the other hand, 
recommends which proof method is likely to be useful
to a given proof goal,
using a supervised learning applied to the 
Archive of Formal Proofs \cite{AFP}.
The key of \pamper{} was its feature extractor:
\pamper{} first applies 108 assertions to each invocation of proof methods
and converts each pair of a proof goal with its context and 
the name of proof method applied to that goal into an array of boolean values
of length 108
because this simpler format is amenable for machine learning algorithms to analyze.
The limitation of \pamper{} is, unlike \psl{}, 
it cannot recommend which arguments in the \verb|induct| method
to tackle a given proof goal.

Taking the same approach as \pamper{},
we attempted to build a recommendation tool, \meloid{} \cite{meloid}, to automatically suggest
promising arguments for the \verb|induct| method
without completing a proof:
we wrote many assertions in Isabelle/ML.
Unfortunately, 
encoding induction heuristics as assertions
directly in Isabelle/ML caused an immense amount of code-clutter,
and we could not encode 
even the human-friendly notion of depth in syntax tree 
since multi-arity functions are represented as curried functions in Isabelle.
Therefore, we developed \lifter{}, expecting that 
\lifter{} serves as 
a \underline{l}anguage for \underline{f}eature \underline{e}xtraction.

We hope that when combined into the supervised learning framework of \meloid{},
assertions written in \lifter{} extract 
the essence of induction in Isabelle/HOL
in a cross-domain style and produce a useful database for machine learning algorithms,
so that new Isabelle users can have the recommendation of promising arguments
for the \verb|induct| method in a fully automatic way.


%
%
 \bibliographystyle{splncs04}
 \bibliography{bibfile}
%

\end{document}